**Realization of an intrinsic ferromagnetic topological state in $MnBi_8Te_{13}$**

(Dated: Apr 22, 2020)

Chaowei Hu[1+], Lei Ding[2+], Kyle N. Gordon[3], Barun Ghosh[4], Hung-Ju Tien[5], Haoxiang Li[3], A. Garrison Linn[3], Shang-Wei Lian[5], Cheng-Yi Huang[6], Scott Mackey[1], Jinyu Liu[1], P. V. Sreenivasa Reddy[5], Bahadur Singh[7,8], Amit Agarwal[4], Arun Bansil[7], Miao Song[10], Dongsheng Li[10], Su-Yang Xu[9], Hsin Lin[6], Huibo Cao[2], Tay-Rong Chang[5,11,12‡], Dan Dessau[3,13†], Ni Ni[1*]

[1]Department of Physics and Astronomy and California NanoSystems Institute, University of California, Los Angeles, CA 90095, USA

[2]Neutron Scattering Division, Oak Ridge National Laboratory, Oak Ridge, Tennessee 37831, USA

[3]Department of Physics, University of Colorado, Boulder, CO 80309, USA

[4]Department of Physics, Indian Institute of Technology-Kanpur, Kanpur, 208016, India

[5]Department of Physics, National Cheng Kung University, Tainan 701, Taiwan

[6]Institute of Physics, Academia Sinica, Taipei 11529, Taiwan

[7]Department of Physics, Northeastern University, Boston, Massachusetts 02115, USA

[8]SZU-NUS Collaborative Center and International Collaborative Laboratory of 2D Materials for Optoelectronic Science & Technology, Engineering Technology Research Center for 2D Materials Information Functional Devices and Systems of Guangdong Province, College of Optoelectronic Engineering, Shenzhen University, ShenZhen 518060, China

[9]Department of Physics, Massachusetts Institute of Technology, Cambridge, Massachusetts 02139, USA

[10]Physical and Computational Sciences Directorate, Pacific Northwest National Laboratory, Richland, WA, 99352, United States

[11]Center for Quantum Frontiers of Research & Technology (QFort), Tainan 701, Taiwan

[12] Physics Division, National Center for Theoretical Sciences, Hsinchu, Taiwan.

[13] CEQM

* nini@physics.ucla.edu

† dessau@colorado.edu

‡ u32trc00@phys.ncku.edu.tw

+ These authors contributed equally.

**ABSTRACT**

The interplay between topology and magnetism is essential for realizing novel topological states including the axion insulator, the magnetic Weyl semimetal, etc. An intrinsically ferromagnetic topological material with only the topological bands at the charge neutrality energy has so far remained elusive. By rationally designing the natural heterostructure consisting of [$MnBi_2Te_4$] septuple layers and [$Bi_2Te_3$] quintuple layers, we report $MnBi_8Te_{13}$ as the first intrinsic ferromagnetic topological material with clean low-energy band structure. Based on the thermodynamic, transport and neutron diffraction measurements, our data show that despite the adjacent [$MnBi_2Te_4$] being 44.1 Å apart, $MnBi_8Te_{13}$ manifests long-range ferromagnetism below 10.5 K with strong coupling between magnetism and charge carriers. Our first-principles calculations and angle-resolved photoemission spectroscopy measurements further demonstrate that $MnBi_8Te_{13}$ is an intrinsic ferromagnetic axion state. Therefore, $MnBi_8Te_{13}$ serves as an ideal system to investigate rich emergent phenomena, including the quantized anomalous Hall effect and quantized topological magnetoelectric effect.

## Introduction

Over the past decade, topology has taken the center stage in condensed matter physics and materials science, emerging as an organizing principle of the states of matter(*1*). Many topological phases, such as quantum spin Hall insulators, quantum anomalous Hall (QAH) insulators, three-dimensional topological insulators (TI), and Weyl semimetals, have been observed(*2*, *3*). Despite tremendous progress, the dominant majority of known topological materials (apart from the QAH insulators) are nonmagnetic materials whereas studies on magnetic topological materials have been far more limited. In contrast to their nonmagnetic counterparts, magnetic topological materials allow for a distinct set of new topological states including the axion insulator, the magnetic Weyl semimetals, the Chern insulators, as well as the three-dimensional QAH insulators(*2*, *3*). Furthermore, magnetism is a natural way to induce non-negligible electronic interactions, paving the way for studying the interplay between band topology and correlations. Therefore, topological magnetic materials have emerged as the frontier of the field. Indeed, since the realization of the QAH state in the magnetically-doped topological insulator thin films $Cr_{0.15}(Bi_{0.1}Sb_{0.9})_{1.85}Te_3$(*4*), recent studies have identified a range of topological magnetic materials such as $Fe_3Sn_2$, $Co_3Sn_2S_2$, $Mn_3Ge$ and $Co_2MnGa$(*5*–*9*). However, the magnetically-doped topological insulator thin films are nonstoichiometric systems where disorder and inhomogeneity are unavoidable. $Fe_3Sn_2$, $Co_3Sn_2S_2$, $Mn_3Ge$ and $Co_2MnGa$ are all large carrier density metals where significant topologically trivial bands coexist with the topological bands at the chemical potential (*5*–*9*). To avoid these drawbacks, an intrinsically ferromagnetic (FM) topological material with magnetic ions occupying their own crystallographic sites and with only the topological bands at the charge neutrality energy is strongly desired but has so far remained elusive.

Recently, $MnBi_2Te_4$, a van der Waals (vdW) compound composed of the septuple layers (SL) of [$MnBi_2Te_4$] was identified as an intrinsic magnetic topological material with clean band structure(*10*–*32*). Unfortunately, magnetic moments in $MnBi_2Te_4$ are antiferromagnetically coupled (AFM) across adjacent [$MnBi_2Te_4$] planes(*15*, *31*). Therefore, even though the quantized anomalous Hall conductance was observed at a record-breaking temperature of 4.5 K, it requires an external magnetic field as large as 12 T to polarize the system into the FM state(*21*, *22*). Can we reduce the interlayer AFM coupling between the adjacent Mn layers to realize intrinsic FM? One material design strategy is to increase the interlayer distance between the adjacent Mn layers. How can we achieve it? Structurally, SL blocks have great compatibility with quintuple (QL) blocks of [$Bi_2Te_3$], as suggested by the existence of $GeBi_4Te_7$ which has alternating [$GeBi_2Te_4$] SL and [$Bi_2Te_3$] QL building blocks(*33*). This superior compatibility provides us flexible structural control to reduce the interlayer magnetic coupling by increasing the interlayer distance between the adjacent [$MnBi_2Te_4$] layers. Based on our material design strategy, here we report the discovery of a novel intrinsic FM TI $MnBi_8Te_{13}$. Despite the interlayer distance between the adjacent [$MnBi_2Te_4$] SLs being 44.1 Å, it is striking and surprising that our thermodynamic, transport and neutron diffraction measurements indicate that $MnBi_8Te_{13}$ has long-range FM order below 10.5 K with the easy axis along the *c* axis. Our first-principles calculations and angle-resolved photoemission spectroscopy (ARPES) measurements further suggest it is an intrinsic FM axion state. Considering the natural heterostructure nature of $MnBi_8Te_{13}$, our finding provides a superior material realization to explore zero-field QAH effect, quantized topological magnetoelectric effect and associated phenomena.

## Results

**Crystal structure of $MnBi_8Te_{13}$.** Although the existence of $MnBi_8Te_{13}$ was mentioned(*34*), its crystal structure was never reported, partially due to the difficulty in growing the $MnBi_8Te_{13}$ phase and separating it from the other members in the $MnBi_{2n}Te_{3n+1}$ family. We managed to grow high quality $MnBi_8Te_{13}$ single crystals and solved the crystal structure of $MnBi_8Te_{13}$ by refining the room-temperature powder X-ray diffraction (XRD) pattern using various structure models. We found that $MnBi_8Te_{13}$ crystallizes in the $R-3m$ symmetry with the lattice parameters: $a = b = 4.37485(7)$ Å, $c = 132.415(3)$ Å, $\alpha = \beta = 90^\circ$, $\gamma = 120^\circ$. The refinement results and structural parameters are summarized in Table S1 and S2. The powder XRD pattern and the Rietveld refinement are shown in Fig. 1a. The low-angle diffraction peaks can be well indexed as the (0 0 L) with lattice parameter *c* as 132.415(3) Å (see the inset of Fig. 1a), which is longer than the parameter *c* of 101.825(8) Å in $MnBi_6Te_{10}$. The formation of $MnBi_8Te_{13}$ is also visualized by the scanning transmission electron microscopy (STEM) image, as shown in Fig. 1b. The STEM image shows vdW structures, which is composed of repeating units of one SL block made of 7 atomic layers and three consecutive QL blocks made of 5 atomic layers. The chemical analysis via the wavelength dispersive spectroscopy measurement results in the elemental composition of the sample as Mn : Bi : Te = 0.74(3) : 8.2(1) : 13, where the Mn/Bi ratio is slightly smaller than the ideal 1/8. The smaller than ideal Mn/Bi ratio is also observed in other $MnBi_{2n}Te_{3n+1}$ due to the Bi substitution on Mn Sites(*16*, *31*).

The crystal structure is shown in Fig. 1c. It is characterized by the alternating stacking of monolayer of $MnTe_6$ octahedra that are well separated by a number of monolayers of $BiTe_6$ octahedra running along the *c*-axis. The distance between the nearest Mn-Mn interlayers is 44.1 Å, which is much larger than 13.6 Å for $MnBi_2Te_4$, 23.8 Å for $MnBi_4Te_7$ and 33.9 Å for $MnBi_6Te_{10}$, as shown in the inset of Fig. 1c. The stacking sequence of the $MnBi_{2n}Te_{3n+1}$ series can be rationalized.(*31*) Using $Bi_2Te_3$ shown in Fig. 1c as the starting point, along the *c* axis, the stacking sequence of $Bi_2Te_3$ is -A-B-C-A-B-C-, where A, B and C represents the bilayers of $BiTe_6$ octahedra whose bottom Te atoms, center Bi atoms and top Te atoms are on the cell edges, respectively. "Mn" layer can replace "A" or "B" or "C" bilayers of $BiTe_6$ octahedra to make $MnBi_{2n}Te_{3n+1}$. For example, as shown in Fig. 1c, for $MnBi_2Te_4$, the stacking sequence is -Mn(B)-C-Mn(A)-B-Mn(C)-A-; for $MnBi_4T_7$, the stacking sequence is -Mn(B)-C-A-; for $MnBi_6T_{10}$, the stacking sequence is -Mn(B)-C-A-B-Mn(C)-A-B-C-Mn(A)-B-C-A-. When it comes to $MnBi_8T_{13}$, the stacking sequence is Mn(B)-C-A-B-C-Mn(A)-B-C-A-B-Mn(C)-A-B-C-A-, exactly identical to the one we obtained based on our powder X-ray refinement. Therefore, with the stacking rule, we can easily assign the stacking sequence for the yet-to-be-discovered higher *n* members of $MnBi_{2n}Te_{3n+1}$ or design new magnetic topological insulators with the QL and SL building blocks.

**Long-range ferromagnetism and its strong coupling with charge carriers in $MnBi_8Te_{13}$.** As a comparison, the physical properties of $MnBi_6Te_{10}$ where two [$Bi_2Te_3$] QL are sandwiched between the adjacent [$MnBi_2Te_4$] SL are presented too. The inset of Figs. 2(a, b) present the temperature dependent specific heat data of $MnBi_8Te_{13}$ and $MnBi_6Te_{10}$, respectively. A specific heat anomaly associated with the magnetic phase transition is observed, which determines the ordering temperature as 10.5 K for $MnBi_8Te_{13}$ and 11.0 K for $MnBi_6Te_{10}$. The magnetic properties of $MnBi_8Te_{13}$ are shown in Figs. 2(a, c, e) whereas those of $MnBi_6Te_{10}$ are presented in Figs. 2(b, d, f). The data indicate that with the *c* axis as the easy axis, $MnBi_8Te_{13}$ is FM below 10.5 K while $MnBi_6Te_{10}$ is AFM below 11.0 K. Figures 2(a,b) present the zero-field-cooled (ZFC) and field-cooled (FC) magnetic susceptibility data, $\chi^c$ ($H \parallel c$) and $\chi^{ab}$ ($H \parallel ab$), measured at 0.1 kOe for $MnBi_8Te_{13}$ and $MnBi_6Te_{10}$, respectively. Sharp contrast can be seen. A large bifurcation of ZFC and FC data of $\chi^c$ appears below 10.5 K in $MnBi_8Te_{13}$, where upon cooling the ZFC data decrease but the FC data increase, suggesting the formation of FM domains. However, for $MnBi_6Te_{10}$, we observed a sharp cusp feature centering at 11.0 K in $\chi^c$, similar to the ones in AFM $MnBi_2Te_4$ and $MnBi_4Te_7$(*17*, *35*) but with a small bifurcation of ZFC and FC data below 9 K. Furthermore, at 2 K, the magnitude of the FC $\chi^c$ in $MnBi_8Te_{13}$ is orders larger than that in AFM $MnBi_2Te_4$, $MnBi_4Te_7$ and $MnBi_6Te_{10}$(*17*, *35*). This strongly suggests different types of ground states in these two materials with $MnBi_8Te_{13}$ being FM and $MnBi_6Te_{10}$ being AFM. Our conclusion is further confirmed by the hysteresis loop of isothermal magnetization curves $M^c(H)$ ($H \parallel c$) shown in Fig. 2(c, d) for $MnBi_8Te_{13}$ and $MnBi_6Te_{10}$, respectively. At 2 K, unlike the $M^c(H)$ data where multiple-step features are observed due to the spin-flop transition in $MnBi_2Te_4$(*15*, *17*) and spin-flip transition in $MnBi_4Te_7$(*35*) and $MnBi_6Te_{10}$, $M^c(H)$ in $MnBi_8Te_{13}$ shows a typical hysteresis loop for FM materials with the coercivity of $H_c = 0.75$ kOe and saturation remanence of $M_r = 3.1\ \mu_B$/Mn. Upon warming, $H_c$ decreases as the hysteresis loop shrinks in area due to the enhanced thermal fluctuations. Figure 2e shows the $M^{ab}(H)$ and $M^c(H)$ of $MnBi_8Te_{13}$ up to 7 T, where a field of 10.4 kOe is required to force spins to saturate in the *ab* plane. This value is 11.3 kOe for $MnBi_6Te_{10}$ (Figs. 2f). Both sets of data indicate Ising anisotropy with the *c* axis as the easy axis in these two compounds. Ising anisotropy has also been suggested in $MnBi_2Te_4$ by the inelastic neutron scattering measurement(*28*). The saturation moment measured at 7 T for $MnBi_8Te_{13}$ is $M_s$ = 3.5(1) $\mu_B$/Mn whereas it is 3.4(1) $\mu_B$/Mn for $MnBi_6Te_{10}$. In both cases, $M_s$ is smaller than 4.6$\mu_B$/Mn, the obtained value from our DFT calculation, but similar to 3.6 $\mu_B$/Mn in $MnBi_2Te_4$(*17*) and 3.5 $\mu_B$/Mn in $MnBi_4Te_7$ (*35*). The reduced Mn saturation moment in $MnBi_8Te_{13}$ is likely due to the Mn site disorder as suggested in $MnBi_2Te_4$ and $MnBi_4Te_7$ (*16*)[,34]. The fitted Weiss temperatures of $MnBi_8Te_{13}$ are $\Theta_w^{ab}$ = 12.8 K and $\Theta_w^{c}$ = 12.1 K, suggesting FM exchange interactions. The fitted effective moments are $\mu_{eff}^{ab} = 5.4\mu_B$/Mn and $\mu_{eff}^{c} = 5.1\mu_B$/Mn, indicating weak single-ion anisotropy. Although the effective moment is smaller than 5.9 $\mu_B$/Mn for a $Mn^{2+}$ ion, it is similar to $MnBi_2Te_4$, $MnBi_4Te_7$ and $MnBi_6Te_{10}$(*17*, *35*).

Figure 2g presents the temperature-dependent in-plane ($\rho_{xx}$) and out-of-plane resistivity ($\rho_{zz}$) of $MnBi_8Te_{13}$. Above 20 K, upon cooling, both $\rho_{xx}$ and $\rho_{zz}$ decrease semi-linearly. Upon further cooling, when approaching $T_c$, due to the stronger scattering from enhanced spin fluctuations, both values increase(*36*, *37*). Then a sharp drop appears in both curves below 10.5 K due to the loss of spin scattering. The drop in $\rho_{zz}$ below $T_c$ is distinct from that in $MnBi_2Te_4$, $MnBi_4Te_7$ and $MnBi_6Te_{10}$ (see Fig. 2h)(*17*, *35*) where the antiparallelly aligned Mn moments enhance resistivity via the spin-flip scattering. Therefore, the drop in $\rho_{zz}$ again suggests FM ordering in $MnBi_8Te_{13}$ since the parallelly aligned Mn moments along the *c* axis eliminate such spin-flip scattering and thus reduce electric resistivity.

To confirm the long-range FM phase transition in $MnBi_8Te_{13}$, we have performed single crystal neutron diffraction experiments at various temperatures. The quality of the instrument resolution limited magnetic Bragg peaks indicates the stacking faults in this crystal are minimal. Figure 3a shows the Bragg reflection (1 0 1) at 4.5 K and 15 K. It is evident that the peak intensity enhances significantly from 15 K to 4.5 K, indicating the development of long-range magnetic ordering with the propagation vector **k**=**0**. Starting from the nuclear space group *R-3m* with Mn atoms occupying the Wyckoff position (0, 0, 0) and **k**=**0**, through symmetry analysis(*38*), we obtained the *k*-maximal magnetic subgroup as $R\text{-}3\,m'$ whose symmetry only allows the presence of a FM arrangement with the Mn spins along the *c* axis. Its magnetic structure is illustrated in Fig. 1c. The order parameter of $MnBi_8Te_{13}$ was measured from 4.5 to 15 K. As shown in Fig. 3b, a power law was employed to fit the intensity of magnetic reflection (1 0 1) as a function of temperature. The fit yields the critical temperature 9.8(3) K and a critical exponent $\beta$ = 0.4(1). The former is consistent with our thermodynamic measurements. The latter is considerably larger than the value of 0.125 expected for the Ising model in two dimensions, excluding this possibility(*39*). More data at lower temperatures could better define the critical behavior.

Strong coupling between charge carriers and magnetism is observed in $MnBi_8Te_{13}$ through the magnetotransport measurements, as shown in Fig. 4. $\rho_{xx}(H)$, $\rho_{zz}(H)$ and $\rho_{xy}(H)$ follow the same hysteresis as that in *M*(*H*). Using $n = H/e\rho_{xy}$, our 50 K data correspond to an electron carrier density of $1.66 \times 10^{20}$ $cm^{-3}$, similar to that of $MnBi_2Te_4$(*26*). Figure 4f shows the $\rho_{xx}(H)$ curves up to 9 T at various temperatures. The overall "W" shape was observed at low temperatures. The "W"-shaped *MR*(*H*) was previously observed in $MnBi_4Te_7$[38], where it was suggested to be a combination of non-magnetic parabolic MR contribution and negative MR originating from FM fluctuations. Unlike $MnBi_4Te_7$ where the "W"-shaped behavior was observed far above its transition temperature 12 K, in $MnBi_8Te_{13}$, the MR quickly changes from the deepest "W" shape at $T_c$ into a parabolic shape at just a few degrees above it. This may suggest weak FM fluctuations above $T_c$ in $MnBi_8Te_{13}$.

**Ferromagnetic axion state revealed by DFT calculation.** To investigate the band topology, in the FM configuration with the spin oriented along the *c* axis, we calculated the electronic band structure of $MnBi_8Te_{13}$. Using the experimentally determined lattice parameters and atomic positions, our calculation shows only a continuous bulk energy gap (Supplementary Note III). In contrast, using the experimental lattice parameters with relaxed atomic positions, our calculation indicates a 170 meV global energy band gap (Fig. 5b). By comparing our DFT calculation with the experimental ARPES data, the one with relaxed atomic positions describes the material well.

To highlight the spin-splitting in the presence of FM ordering, we present the $<S_z>$ resolved band-structure in Fig. 5c. The band structure projected on the Bi *p* and Te *p* orbitals shows that the bands near the Fermi level mostly originate from the Bi *p* and Te *p* orbitals. As shown in Fig. 5d, there are clear band inversions between the Bi $p_z$ and Te $p_z$ states. In fact, the Bi $p_z$ orbitals reach deep into the valence bands, indicating multiple possible band inversions that originate from the different [$Bi_2Te_3$] QL and [$MnBi_2Te_4$] SL of $MnBi_8Te_{13}$.

The presence of clear band inversions around the Γ point hints towards a topological phase. To unravel the exact topology of this system, we first compute the Chern number in the $k_z = 0$ and $k_z = \pi$ planes. In both planes the Chern number is found to be zero. Next, we compute the parity-based higher-order $Z_4$ invariant, which is given by

$$Z_4 = \sum_{i=1}^{8} \sum_{n=1}^{n_{occ}} \frac{1+\xi_n(\Gamma_i)}{2} mod 4. \quad (1)$$

Here, $\xi_n(\Gamma_i)$ is the parity of the $n^{th}$ band at the $i^{th}$ time reversal invariant momentum (TRIM) point $\Gamma_i$, and $n_{occ}$ is the number of occupied bands. The $Z_4$ invariant is well defined for an inversion symmetric system, even in the absence of time reversal symmetry(*40–42*). The odd values of $Z_4$ (1, 3) indicate a Weyl semimetal phase, while $Z_4 =2$ implies an insulator phase with a quantized topological magnetoelectric effect (axion coupling $\theta = \pi$)(*43*). A detailed list of the number of occupied bands with even ($n^{+}_{occ}$) and odd ($n^{-}_{occ}$) parity eigenvalues at the eight TRIM points are shown in Supplementary Table S3. Based on this, the computed $Z_4$ invariant is found to be 2, which demonstrates that $MnBi_8Te_{13}$ is an intrinsic FM axion insulator.

It is noted that, in calculations with or without atomic relaxation, the characteristics of band inversion and the topology of whole system remain the same, i.e., an FM axion state. In addition, we have investigated the evolution of the band structures by changing lattice constants *a*, *b* (in-plane) and *c* (out-of-plane) simultaneously. Within the range from -3% to +3% in $MnBi_8Te_{13}$, we did not find any crossing point or the additional band inversion feature. Our calculation reveals that the axion phase in $MnBi_8Te_{13}$ is quite stable.

**Surface state revealed by ARPES and DFT** To investigate if surface states appear in $MnBi_8Te_{13}$, we have performed small-spot (20 μm × 50 μm) ARPES scanned across the surfaces of $MnBi_8Te_{13}$. According to the crystal structure of $MnBi_8Te_{13}$, four different terminations are expected during the cleave, as visualized in the cartoon pictures in Fig. 6. Our ARPES data and DFT calculations indeed reveal distinguishing surface states for the four different terminations of $MnBi_8Te_{13}$, which are summarized in Fig. 6.

Figures 6(a-d) show the isoenergy surfaces at the Fermi level which uniquely fingerprints each of the four possible terminations of $MnBi_8Te_{13}$. We identify a unique circular Fermi surface for the SL termination which has been observed in other $MnBi_{2n}Te_{3n+1}$ ($n$ = 1, 2, 3) (*44–46*). We find that the $QL_n$ terminations show a dominant six-fold symmetric Fermi surface with decreasing cross-sectional area from $QL_1$ to $QL_3$. Our assignment of the isoenergy surfaces to the respective terminations for $MnBi_8Te_{13}$ is consistent with previous measurements on the simpler $MnBi_2Te_4$, $MnBi_4Te_7$, and $MnBi_6Te_{10}$ compounds(*24*, *44–*

*46*). Figures 6(e-h) present the experimental ARPES *E-k* maps along the $M \rightarrow \Gamma \rightarrow M$ high symmetry direction. Different types of surface states are observed for each of the four terminations. For the SL termination, a gapless or nearly gapless surface state appears (Fig. 6e), consistent with the gapless surface state observed on the SL termination in previous ARPES measurements of $MnBi_2Te_4$, $MnBi_4Te_7$, and $MnBi_6Te_1$(*24*, *44–46*). The $QL_3$ termination, which is unique to $MnBi_8Te_{13}$, shows mass renormalization near the Fermi level (Fig. 6h). Furthermore, a large surface gap of ~105 meV centering around the charge neutrality point of -0.35 eV is clearly revealed in the $QL_1$ termination and is highlighted by the arrow (Fig. 6f). It is noted, we find that the spectra taken below and above the transition temperature are very similar in terms of the size of the gap, except for thermal broadening of the electrons (Supplementary Note II). This has been observed in various magnetic topological insulators, including $MnBi_2Te_4$, $MnBi_4Te_7$, $MnBi_6Te_{10}$ and the origin of it is under debate(*24*, *47*). This aspect will also be discussed more in Supplementary Note II.

In order to confirm the topological nature of $MnBi_8Te_{13}$, we calculated the surface spectral weight throughout the (001) surface BZ using the semi-infinite Green's function approach for SL, $QL_1$, $QL_2$, and $QL_3$ terminations correspondingly, as shown in Fig. 6(i-l). Each state is plotted with a color corresponding to the integrated charge density of the state within the topmost QL or SL. In addition, we shift the Fermi level to match the experimentally observed Fermi level.

Excellent agreement between the experimental ARPES data and DFT calculation is achieved for the $QL_n$ terminations as shown in Fig. 6. Our calculation suggests that the sizable surface band gap presented in the $QL_1$ termination around -0.35 eV is a hybridization gap induced from the hybridization effect between the topmost QL and the nearest-neighboring SL (Supplementary Note IV). Although the calculated band gap value is smaller than that from ARPES, it is a well-known problem in that GGA generally underestimates the band gap in semiconductors and insulators(*48*, *49*).The hybridization gap seems universal in $MnBi_{2n}Te_{3n+1}$, which has been suggested for $MnBi_6Te_{10}$(*50*) and $MnBi_4Te_7$(*51*). However, until now, it is unclear if the hybridization gap exists and supports QAH in the 2D limit. To settle down this issue, we have performed DFT calculation on a 7-layered finite-size slab model with the arrangement of vacuum-[QL-SL-QL-QL-QL-SL-QL]-vacuum, i.e. a structure that has the $QL_1$ termination on both surfaces. The main features of the slab results shown in Fig. 5e are consistent with our semi-infinite Green's function approach. We then calculate the Chern number of this slab model based on the Wilson loop method(*52*). Our result shows a nontrivial Chern number ($C$ = 1) in this hybridization gap (Fig. 5f), demonstrating that $MnBi_8Te_{13}$ is indeed a QAH insulator in its 2D limit if the Fermi level is gated to the middle of the large hybridization gap.

As a consequence of strong exchange fields from the Mn magnetic layer in the SL termination, our calculation on the SL termination results in a parabolic large gapped surface band dispersion in the bulk energy gap (Fig. 6i). This is in sharp contrast with the gapless Dirac surface state revealed by the ARPES (Fig. 6e). For the SL termination, the deviation between ARPES and DFT calculation as well as the gapless Dirac surface state are universal in the $MnBi_{2n}Te_{3n+1}$ family ($n$ = 1, 2, 3)(*24*, *44–46*). The unexpected gapless surface state is argued to be caused by the surface spin reconstruction when the magnetic [$MnBi_2Te_4$] layer is at the vacuum-sample interface(*24*, *45*, *46*). Further experiments in identifying the spin reconstruction on the SL termination are urged to settle down this issue.

## Discussion

We have presented the realization of the intrinsic ferromagnetic topological material, $MnBi_8Te_{13}$. Our work has several implications. First, our theoretical calculations show that $MnBi_8Te_{13}$ is a ferromagnetic axion insulator. Such a topological axion state suggests a quantized magnetoelectric coupling and an emergent axion electrodynamics. Therefore, the optical responses of $MnBi_8Te_{13}$, especially in the terahertz regime, may be of great interest. Second, the intrinsic ferromagnetism paves the way for the realization of QAH state at zero magnetic field. Third, when it is exfoliated into the 2D version, the superlattice nature and the stacking of vdW $MnBi_8Te_{13}$ makes it possible to fabricate richer combinations of natural heterostructures than those from AFM $MnBi_{2n}Te_{3n+1}$ ($n$ = 1, 2, 3). Various emergent properties such as QAH state and QSH state are proposed(*36*) for such heterostructures. Furthermore, the observation of long-range magnetic order is striking considering the extremely large separation of 44.1 Å between the adjacent [$MnBi_2Te_4$] SLs; further investigations will shed light on the mechanism of the long-range magnetic ordering in $MnBi_8Te_{13}$ and will advance our understanding on the magnetism in vdW materials. Finally, our work here in general establishes natural heterostructuring as a powerful way to rationally design and control magnetism and other broken symmetry states in layered vdW materials.

## Methods

**Sample growth and characterization.** We have grown single crystals of $MnBi_{2n}Te_{3n+1}$ ($n$ = 3 and 4) using self-flux(*17*, *35*). Around 20g of Mn, Bi and Te elements with the molar ratio of MnTe: $Bi_2Te_3$ as 15:85 are loaded into a 5-mL alumina crucible and then sealed inside a quartz tube under 1/3 atmosphere of Ar. The sample ampule was heated up to 900 $^{o}$C, stayed for 5 hours, then quickly air quenched to room temperature. It was then placed inside a furnace which was preheated to 595 $^{o}$C. The ampule stayed at 595 $^{o}$C for 5 hours and then slowly cooled down to the decanting temperature in 48 hours. We then let the ampule stay at the decanting temperature for about 24 hours before we separated the single crystals from the liquid flux using a centrifuge. By varying the decanting temperature, we could obtain different phases of $MnBi_{2n}Te_{3n+1}$ (Supplementary Note I). We found out that the growth window for $MnBi_4Te_7$ and $MnBi_6Te_{10}$ are around two degrees while the one for $MnBi_8Te_{13}$ is limited to be only one degree barely above the melting temperature of the flux. The method to determine the decanting temperature for each phase is described in Supplementary Note I in details. $Bi_2Te_3$ is the inevitable side product. We also noticed that with increasing $n$ numbers, the chance of the intergrowth between $Bi_2Te_3$ and $MnBi_{2n}Te_{3n+1}$ ($n$ = 3, 4) got enhanced. Therefore, extra care was paid in screening out the right piece. X-ray diffraction at low angles for both the top and bottom (0 0 $l$) surfaces as well as powder X-ray diffraction were performed on a PANalytical Empyrean diffractometer (Cu Ka radiation). Structural determination based on powder X-ray diffraction was done using FullProf Suite software(*53*). Scanning transmission electron microscopy was measured on a sample prepared by focused-ion beam (FIB) and images are made on a Cs Corrected transmission electron microscope (Titan, Thermo Fisher, FEI) at 300 kV. Images are Bragg filtered. Electric resistivity was measured in a Quantum Design (QD) DynaCool Physical Properties Measurement System (DynaCool PPMS). All samples were shaped into thin rectangular bars and the four- and six-probe configurations were used for electrical resistivity and Hall resistivity, respectively. The magnetoresistivity was symmetrized using $\rho_{xx}(B)=(\rho_{xx}(B)+\rho_{xx}(-B))/2$ and the Hall resistivity was antisymmetrized using $\rho_{xy}(B)=(\rho_{xy}(B)-\rho_{xy}(-B))/2$. The magnetization was measured using QD Magnetic Properties Measurement System (QD MPMS). The piece used for the magnetization measurement was later ground into fine powder, whose powder XRD pattern shown no impurity and were used for structural determination.

**Single-crystal neutron diffraction** experiments were performed at the HB-3A DEMAND single crystal neutron diffractometer equipped with a two-dimensional scintillation Anger camera with 0.6-mm spatial resolution at the High Flux Isotope Reactor (HFIR) at ORNL in the temperature range of 4.5-15 K(*54*). Neutron wavelength of 1.551 Å was selected by using a bent perfect Si-220 monochromator. Rocking curve scans, which cover the full peak profile of the Bragg reflection (1 0 1), have been carried out at 4.5 K and 15 K, respectively. The peak (1 0 1) was selected for revealing the magnetic ordering parameter from 4.5 K to 15 K.

**First-principles calculations.** The bulk band structures of $MnBi_8Te_{13}$ were computed using the projector augmented wave method as implemented in the VASP package(*55–57*) within the generalized gradient approximation (GGA)(*58*) and GGA plus Hubbard $U$ (GGA+U)(*59*) scheme. On-site $U$ = 5.0 eV was used for Mn $d$-orbitals. The spin−orbit coupling (SOC) was included self-consistently in the calculations of electronic structures with a Monkhorst–Pack *k*-point mesh 5 × 5 × 5. The experimental lattice parameters were employed. The atomic positions were relaxed until the residual forces were less than 0.01 eV/ Å. We used Mn *d*-orbitals, Bi *p*-orbitals and Te *p*-orbitals to construct Wannier functions, without performing the procedure for maximizing localization(*60*).

The surface spectral weight throughout the (001) surface BZ are calculated using the semi-infinite Green's function approach for four distinct surface terminations. We set up four different models to calculate the surface state to compare with the ARPES data, for example, for the SL termination, the semi-infinite Green's function model is set as vacuum-[SL-QL-QL-QL]-[ SL-QL-QL-QL]-infinite; to get that of the $QL_1$ termination, the model is set to be vacuum-[QL-SL-QL-QL]-[ QL-SL-QL-QL]- infinite, etc. This method is valid since the thickness of the QL or SL is about 10 Å, which roughly equals the typical escape depth of photoelectrons.

**ARPES measurements.** ARPES measurements on single crystals of $MnBi_8Te_{13}$ were carried out at the Stanford Synchrotron Research Laboratory (SSRL) beamline 5-2 with photon energies of 26 eV with linear horizontal polarization and a 20 μm × 50 μm beam spot. Single crystal samples were top-posted on the (001) surface, and cleaved in-situ in an ultra-high vacuum better than $4\times10^{-11}$ Torr and a temperature of 15 K.

## References and Notes

1. M. Z. Hasan, C. L. Kane, Colloquium: topological insulators. *Rev. Mod. Phys.* **82**, 3045 (2010).

2. R. S. K. Mong, A. M. Essin, J. E. Moore, Antiferromagnetic topological insulators. *Phys. Rev. B*. **81**, 245209 (2010).

3. Y. Tokura, K. Yasuda, A. Tsukazaki, Magnetic topological insulators. *Nat. Rev. Phys.*, 1 (2019).

4. C.-Z. Chang, J. Zhang, X. Feng, J. Shen, Z. Zhang, M. Guo, K. Li, Y. Ou, P. Wei, L.-L. Wang, others, Experimental observation of the quantum anomalous Hall effect in a magnetic topological insulator. *Science*. **340**, 167–170 (2013).

5. L. Ye, M. Kang, J. Liu, F. Von Cube, C. R. Wicker, T. Suzuki, C. Jozwiak, A. Bostwick, E. Rotenberg, D. C. Bell, others, Massive Dirac fermions in a ferromagnetic kagome metal. *Nature*. **555**, 638 (2018).

6. I. Belopolski, K. Manna, D. S. Sanchez, G. Chang, B. Ernst, J. Yin, S. S. Zhang, T. Cochran, N. Shumiya, H. Zheng, others, Discovery of topological Weyl fermion lines and drumhead surface states in a room temperature magnet. *Science*. **365**, 1278–1281 (2019).

7. A. K. Nayak, J. E. Fischer, Y. Sun, B. Yan, J. Karel, A. C. Komarek, C. Shekhar, N. Kumar, W. Schnelle, J. Kübler, others, Large anomalous Hall effect driven by a nonvanishing Berry curvature in the noncolinear antiferromagnet $Mn_3Ge$. *Sci. Adv.* **2**, e1501870 (2016).

8. E. Liu, Y. Sun, N. Kumar, L. Muechler, A. Sun, L. Jiao, S.-Y. Yang, D. Liu, A. Liang, Q. Xu, others, Giant anomalous Hall effect in a ferromagnetic kagome-lattice semimetal. *Nat. Phys.* **14**, 1125 (2018).

9. Q. Wang, Y. Xu, R. Lou, Z. Liu, M. Li, Y. Huang, D. Shen, H. Weng, S. Wang, H. Lei, Large intrinsic anomalous Hall effect in half-metallic ferromagnet $Co_3Sn_2S_2$ with magnetic Weyl fermions. *Nat. Commun.* **9**, 3681 (2018).

10. T. Hirahara, S. V. Eremeev, T. Shirasawa, Y. Okuyama, T. Kubo, R. Nakanishi, R. Akiyama, A. Takayama, T. Hajiri, S. Ideta, others, Large-gap magnetic topological heterostructure formed by subsurface incorporation of a ferromagnetic layer. *Nano Lett.* **17**, 3493–3500 (2017).

11. J. A. Hagmann, X. Li, S. Chowdhury, S.-N. Dong, S. Rouvimov, S. J. Pookpanratana, K. M. Yu, T. A. Orlova, T. B. Bolin, C. U. Segre, others, Molecular beam epitaxy growth and structure of self-assembled $Bi_2Se_3$/$Bi_2MnSe_4$ multilayer heterostructures. *New J. Phys.* **19**, 085002 (2017).

12. M. M. Otrokov, T. V. Menshchikova, M. G. Vergniory, I. P. Rusinov, A. Y. Vyazovskaya, Y. M. Koroteev, G. Bihlmayer, A. Ernst, P. M. Echenique, A. Arnau, others, Highly-ordered wide bandgap materials for quantized anomalous Hall and magnetoelectric effects. *2D Mater.* **4**, 025082 (2017).

13. Y. Gong, J. Guo, J. Li, K. Zhu, M. Liao, X. Liu, Q. Zhang, L. Gu, L. Tang, X. Feng, others, Experimental realization of an intrinsic magnetic topological insulator. *Chin. Phys. Lett.* **36**, 076801 (2019).

14. D. S. Lee, T.-H. Kim, C.-H. Park, C.-Y. Chung, Y. S. Lim, W.-S. Seo, H.-H. Park, Crystal structure, properties and nanostructuring of a new layered chalcogenide semiconductor, $Bi_2MnTe_4$. *CrystEngComm*. **15**, 5532–5538 (2013).

15. M. Otrokov, I. Klimovskikh, H. Bentmann, D. Estyunin, A. Zeugner, Z. Aliev, S. Gaß, A. Wolter, A. Koroleva, A. Shikin, others, Prediction and observation of an antiferromagnetic topological insulator. *Nature*. **576**, 416–422 (2019).

16. A. Zeugner, F. Nietschke, A. U. Wolter, S. Ga\s s, R. C. Vidal, T. R. Peixoto, D. Pohl, C. Damm, A. Lubk, R. Hentrich, others, Chemical Aspects of the Candidate Antiferromagnetic Topological Insulator $MnBi_2Te_4$. *Chem. Mater.* (2019).

17. J.-Q. Yan, Q. Zhang, T. Heitmann, Z. Huang, K. Chen, J.-G. Cheng, W. Wu, D. Vaknin, B. Sales, R. McQueeney, Crystal growth and magnetic structure of $MnBi_2Te_4$. *Phys. Rev. Mater.* **3**, 064202 (2019).

18. M. Otrokov, I. Rusinov, M. Blanco-Rey, M. Hoffmann, A. Y. Vyazovskaya, S. Eremeev, A. Ernst, P. Echenique, A. Arnau, E. Chulkov, Unique Thickness-Dependent Properties of the van der Waals Interlayer Antiferromagnet $MnBi_2Te_4$ Films. *Phys. Rev. Lett.* **122**, 107202 (2019).

19. D. Zhang, M. Shi, T. Zhu, D. Xing, H. Zhang, J. Wang, Topological Axion States in the Magnetic Insulator $MnBi_2Te_4$ with the Quantized Magnetoelectric Effect. *Phys. Rev. Lett.* **122**, 206401 (2019).

20. J. Li, Y. Li, S. Du, Z. Wang, B.-L. Gu, S. Zhang, K. He, W. Duan, Y. Xu, Intrinsic magnetic topological insulators in van der Waals layered $MnBi_2Te_4$-family materials. *Sci. Adv.* **5**, eaaw5685 (2019).

21. Y. Deng, Y. Yu, M. Z. Shi, Z. Guo, Z. Xu, J. Wang, X. H. Chen, Y. Zhang, Quantum anomalous Hall effect in intrinsic magnetic topological insulator $MnBi_2Te_4$. *Science*. **367**, 895–900 (2020).

22. C. Liu, Y. Wang, H. Li, Y. Wu, Y. Li, J. Li, K. He, Y. Xu, J. Zhang, Y. Wang, Robust axion insulator and Chern insulator phases in a two-dimensional antiferromagnetic topological insulator. *Nat. Mater.*, 1–6 (2020).

23. B. Chen, F. Fei, D. Zhang, B. Zhang, W. Liu, S. Zhang, P. Wang, B. Wei, Y. Zhang, Z. Zuo, others, Intrinsic magnetic topological insulator phases in the Sb doped $MnBi_2Te_4$ bulks and thin flakes. *Nat. Commun.* **10**, 1–8 (2019).

24. Y.-J. Hao, P. Liu, Y. Feng, X.-M. Ma, E. F. Schwier, M. Arita, S. Kumar, C. Hu, R. Lu, M. Zeng, others, Gapless surface Dirac cone in antiferromagnetic topological insulator $MnBi_2Te_4$. *Phys Rev X*. **9**, 041038 (2019).

25. Y. Chen, L. Xu, J. Li, Y. Li, C. Zhang, H. Li, Y. Wu, A. Liang, C. Chen, S. Jung, others, Topological Electronic Structure and Its Temperature Evolution in Antiferromagnetic Topological Insulator $MnBi_2Te_4$. *Phys Rev X*. **9**, 041040 (2019).

26. H. Li, S.-Y. Gao, S.-F. Duan, Y.-F. Xu, K.-J. Zhu, S.-J. Tian, J.-C. Gao, W.-H. Fan, Z.-C. Rao, J.-R. Huang, others, Dirac surface states in intrinsic magnetic topological insulators $EuSn_2As_2$ and $MnBi_2Te_4$. *Phys. Rev. X*. **9**, 041039 (2019).

27. S. Zhang, R. Wang, X. Wang, B. Wei, H. Wang, G. Shi, F. Wang, B. Jia, Y. Ouyang, B. Chen, others, Experimental observation of the gate-controlled reversal of the anomalous Hall effect in the intrinsic magnetic topological insulator $MnBi_2Te_4$ device. *Nano Lett.* (2019).

28. B. Li, J.-Q. Yan, D. M. Pajerowski, E. Gordon, A.-M. Nedić, Y. Sizyuk, L. Ke, P. P. Orth, D. Vaknin, R. J. McQueeney, Competing Magnetic Interactions in the Antiferromagnetic Topological Insulator $MnBi_2Te_4$. *Phys Rev Lett*. **124**, 167204 (2020).

29. H. Li, S. Liu, C. Liu, J. Zhang, Y. Xu, R. Yu, Y. Wu, Y. Zhang, S. Fan, Antiferromagnetic Topological Insulator $MnBi_2Te_4$: Synthesis and Magnetic properties. *Phys. Chem. Chem. Phys.* **22**, 556–563 (2020).

30. K. Chen, B. Wang, J.-Q. Yan, D. Parker, J.-S. Zhou, Y. Uwatoko, J.-G. Cheng, Suppression of the antiferromagnetic metallic state in the pressurized $MnBi_2Te_4$ single crystal. *Phys. Rev. Mater.* **3**, 094201 (2019).

31. L. Ding, C. Hu, F. Ye, E. Feng, N. Ni, H. Cao, Crystal and magnetic structures of magnetic topological insulators $MnBi_2Te_4$ and $MnBi_4Te_7$. *Phys. Rev. B*. **101**, 020412 (2020).

32. R. Vidal, H. Bentmann, T. Peixoto, A. Zeugner, S. Moser, C.-H. Min, S. Schatz, K. Kißner, M. Ünzelmann, C. Fornari, others, Surface states and Rashba-type spin polarization in antiferromagnetic $MnBi_2Te_4$ (0001). *Phys. Rev. B*. **100**, 121104 (2019).

33. S. Muff, F. von Rohr, G. Landolt, B. Slomski, A. Schilling, R. J. Cava, J. Osterwalder, J. H. Dil, Separating the bulk and surface n-to p-type transition in the topological insulator $GeBi_{4-x}Sb_xTe_7$. *Phys. Rev. B*. **88**, 035407 (2013).

34. Z. A. Jahangirli, E. H. Alizade, Z. S. Aliev, M. M. Otrokov, N. A. Ismayilova, S. N. Mammadov, I. R. Amiraslanov, N. T. Mamedov, G. S. Orudjev, M. B. Babanly, others, Electronic structure and dielectric function of Mn-Bi-Te layered compounds. *J. Vac. Sci. Technol. B Nanotechnol. Microelectron. Mater. Process. Meas. Phenom.* **37**, 062910 (2019).

35. C. Hu, K. N. Gordon, P. Liu, J. Liu, X. Zhou, P. Hao, D. Narayan, E. Emmanouilidou, H. Sun, Y. Liu, others, A van der Waals antiferromagnetic topological insulator with weak interlayer magnetic coupling. *Nat. Commun.* **11**, 1–8 (2020).

36. H. Sun, B. Xia, Z. Chen, Y. Zhang, P. Liu, Q. Yao, H. Tang, Y. Zhao, H. Xu, Q. Liu, Rational Design Principles of the Quantum Anomalous Hall Effect in Superlatticelike Magnetic Topological Insulators. *Phys Rev Lett*. **123**, 096401 (2019).

37. M. A. Kassem, Y. Tabata, T. Waki, H. Nakamura, Low-field anomalous magnetic phase in the kagome-lattice shandite $Co_3Sn_2S_2$. *Phys. Rev. B*. **96**, 014429 (2017).

38. J. Perez-Mato, S. Gallego, E. Tasci, L. Elcoro, G. de la Flor, M. Aroyo, Symmetry-based computational tools for magnetic crystallography. *Annu. Rev. Mater. Res.* **45**, 217–248 (2015).

39. M. F. Collins, *Magnetic critical scattering* (Oxford university press, 1989), vol. 4.

40. A. M. Turner, Y. Zhang, R. S. Mong, A. Vishwanath, Quantized response and topology of magnetic insulators with inversion symmetry. *Phys. Rev. B*. **85**, 165120 (2012).

41. S. Ono, H. Watanabe, Unified understanding of symmetry indicators for all internal symmetry classes. *Phys. Rev. B*. **98**, 115150 (2018).

42. H. Watanabe, H. C. Po, A. Vishwanath, Structure and topology of band structures in the 1651 magnetic space groups. *Sci. Adv.* **4**, eaat8685 (2018).

43. Y. Xu, Z. Song, Z. Wang, H. Weng, X. Dai, Higher-Order Topology of the Axion Insulator $EuIn_2As_2$. *Phys. Rev. Lett.* **122**, 256402 (2019).

44. K. Gordon *et al.* https://arxiv.org/abs/1910.13943 (2019).

45. L. Xu *et al.* https://arxiv.org/abs/1910.11014 (2019).

46. S. Tian *et al.* https://arxiv.org/abs/1910.10101 (2019).

47. N. H. Jo *et al.* https://arxiv.org/abs/1910.14626 (2019).

48. X. Qian, J. Liu, L. Fu, J. Li, Quantum spin Hall effect in two-dimensional transition metal dichalcogenides. *Science*. **346**, 1344–1347 (2014).

49. Z. Fei, T. Palomaki, S. Wu, W. Zhao, X. Cai, B. Sun, P. Nguyen, J. Finney, X. Xu, D. H. Cobden, Edge conduction in monolayer $WTe_2$. *Nat. Phys.* **13**, 677–682 (2017).

50. X.-M. Ma *et al.* https://arxiv.org/abs/1912.13237 (2019).

51. X. Wu *et al.* https://arxiv.org/abs/2002.00320 (2020).

52. R. Yu, X. L. Qi, A. Bernevig, Z. Fang, X. Dai, Equivalent expression of $Z_2$ topological invariant for band insulators using the non-Abelian Berry connection. *Phys Rev B*. **84**, 075119 (2011).

53. J. Rodríguez-Carvajal, Recent advances in magnetic structure determination by neutron powder diffraction. *Phys. B Condens. Matter*. **192**, 55–69 (1993).

54. H. Cao, B. Chakoumakos, K. Andrews, Y. Wu, R. Riedel, J. Hodges, W. Zhou, R. Gregory, B. Haberl, J. Molaison, others, DEMAND, a Dimensional Extreme Magnetic Neutron Diffractometer at the High Flux Isotope Reactor. *Crystals*. **9**, 5 (2019).

55. W. Kohn, L. J. Sham, Self-consistent equations including exchange and correlation effects. *Phys. Rev.* **140**, A1133 (1965).

56. G. Kresse, J. Furthmüller, Efficiency of ab-initio total energy calculations for metals and semiconductors using a plane-wave basis set. *Comput. Mater. Sci.* **6**, 15–50 (1996).

57. G. Kresse, D. Joubert, From ultrasoft pseudopotentials to the projector augmented-wave method. *Phys. Rev. B*. **59**, 1758 (1999).

58. J. P. Perdew, K. Burke, M. Ernzerhof, Generalized Gradient Approximation Made Simple. *Phys Rev Lett*. **77**, 3865–3868 (1996).

59. S. Dudarev, G. Botton, S. Savrasov, C. Humphreys, A. Sutton, Electron-energy-loss spectra and the structural stability of nickel oxide: An LSDA+ U study. *Phys. Rev. B*. **57**, 1505 (1998).

60. N. Marzari, D. Vanderbilt, Maximally localized generalized Wannier functions for composite energy bands. *Phys. Rev. B*. **56**, 12847 (1997).

**Acknowledgements**

Work at UCLA was supported by the U.S. Department of Energy (DOE), Office of Science, Office of Basic Energy Sciences under Award Number DE-SC0011978. Work at CU Boulder was supported by NSF-DMR 1534734. Work at ORNL was supported by the U.S. DOE BES Early Career Award KC0402010 under Contract DEAC05-00OR22725. This research used resources at the High Flux Isotope Reactor, DOE Office of Science User Facilities operated by the ORNL. We acknowledge Makoto Hashimoto and Dong-Hui Lu for help with the ARPES measurements. Use of the Stanford Synchrotron Radiation Lightsource, SLAC National Accelerator Laboratory, is supported by the U.S. Department of Energy, Office of Science, Office of Basic Energy Sciences under Contract No. DE-AC02-76SF00515. BG wants to acknowledge CSIR for the senior research fellowship. BG and AA thank CC-IITK for providing the HPC facility. T.-R.C. was supported by the Young Scholar Fellowship Program from the Ministry of Science and Technology (MOST) in Taiwan, under a MOST grant for the Columbus Program MOST108-2636- M-006-002, National Cheng Kung University, Taiwan, and National Center for Theoretical Sciences, Taiwan. This work was supported partially by the MOST, Taiwan, Grant MOST107-2627-E-006-001.This research was supported in part by Higher Education Sprout Project, Ministry of Education to the Headquarters of University Advancement at National Cheng Kung University (NCKU). TEM characterization and analysis were supported by the U.S. Department of Energy (DOE), Office of Science, Office of Basic Energy Sciences (BES), Division of Materials Science and Engineering (DMSE) through Early Career Research Program under award KC0203020:67037.

**Author contributions**

N. N. conceived the idea and organized the research. C. H., N. N. and S. M. grew the bulk single crystals and carried out X-ray, thermodynamic and transport measurements and data analysis. H. C. and L. D. carried out structure determination and neutron diffraction measurements. K. G., H. L., A. L. and D. D. carried out the ARPES measurements and data analysis. B. G., S. L, H. T., C. H. P. R, B.S., A. A., A. B., T.-R. C., S. X. and L. L. performed the first-principles calculations. M. S. and D. L performed TEM measurements. All authors wrote the manuscript.

**Competing interests**

The authors declare that they have no competing interests.

**Data and materials availability statement**

All data needed to evaluate the conclusions in the paper are present in the paper and/or the Supplementary Materials. Additional data related to this paper may be requested from the authors.

Figure 1

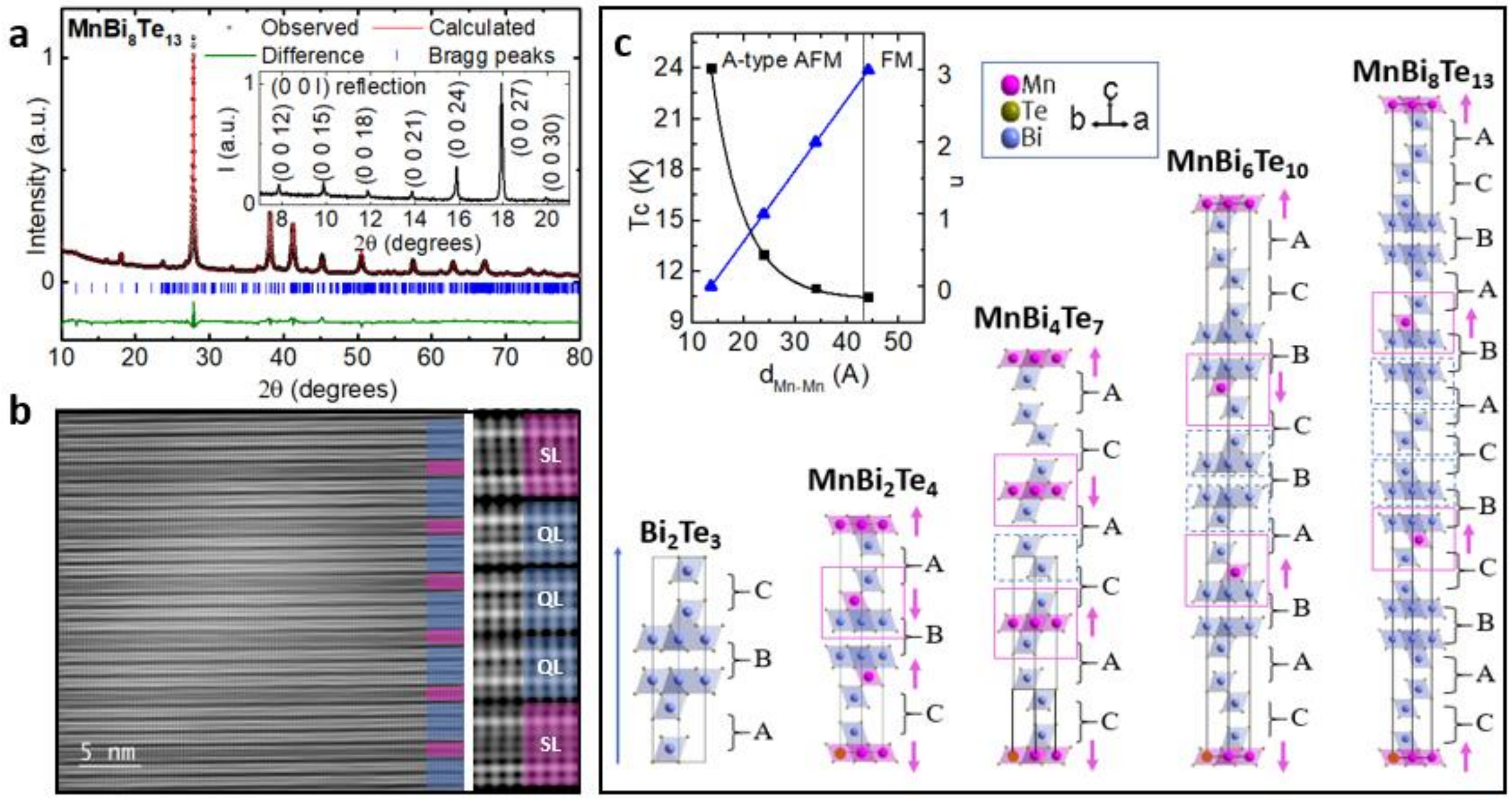

**Figure 1. Crystal structure of MnBi$_8$Te$_{13}$.** **a** Powder X-ray diffraction and the refinement of MnBi$_8$Te$_{13}$. Inset: The low-angle (0 0 *l*) x-ray diffraction peaks of the cleaved *ab* surface of MnBi$_8$Te$_{13}$. **b** STEM images of MnBi$_8$Te$_{13}$ made on a FIB sample. The purple blocks label the [MnBi$_2$Te$_4$] SL and the blue blocks mark the [Bi$_2$Te$_3$] QL. **c** Schematic drawing of the crystal and magnetic structure of MnBi$_{2n}$Te$_{3n+1}$ ($n$ = 0, 1, 2, 3 and 4) with the stacking sequence listed. A, B and C represents the bilayers of BiTe$_6$ octahedra whose bottom Te atoms, center Bi atoms and top Te atoms are on the cell edges, respectively. Magenta arrow: magnetic structure in the order state. Blue block: edge-sharing BiTe$_6$ octahedra; Magenta block: edge-sharing MnTe$_6$ octahedra, which are connected to the blue block via edge-sharing. Inset: *Tc* (The critical temperatures) vs. $d_{Mn-Mn}$ (the interlayer distance between the adjacent Mn-Mn layers) and $n$ vs. $d_{Mn-Mn}$. in MnBi$_{2n}$Te$_{3n+1}$ ($n$ = 1, 2, 3 and 4).

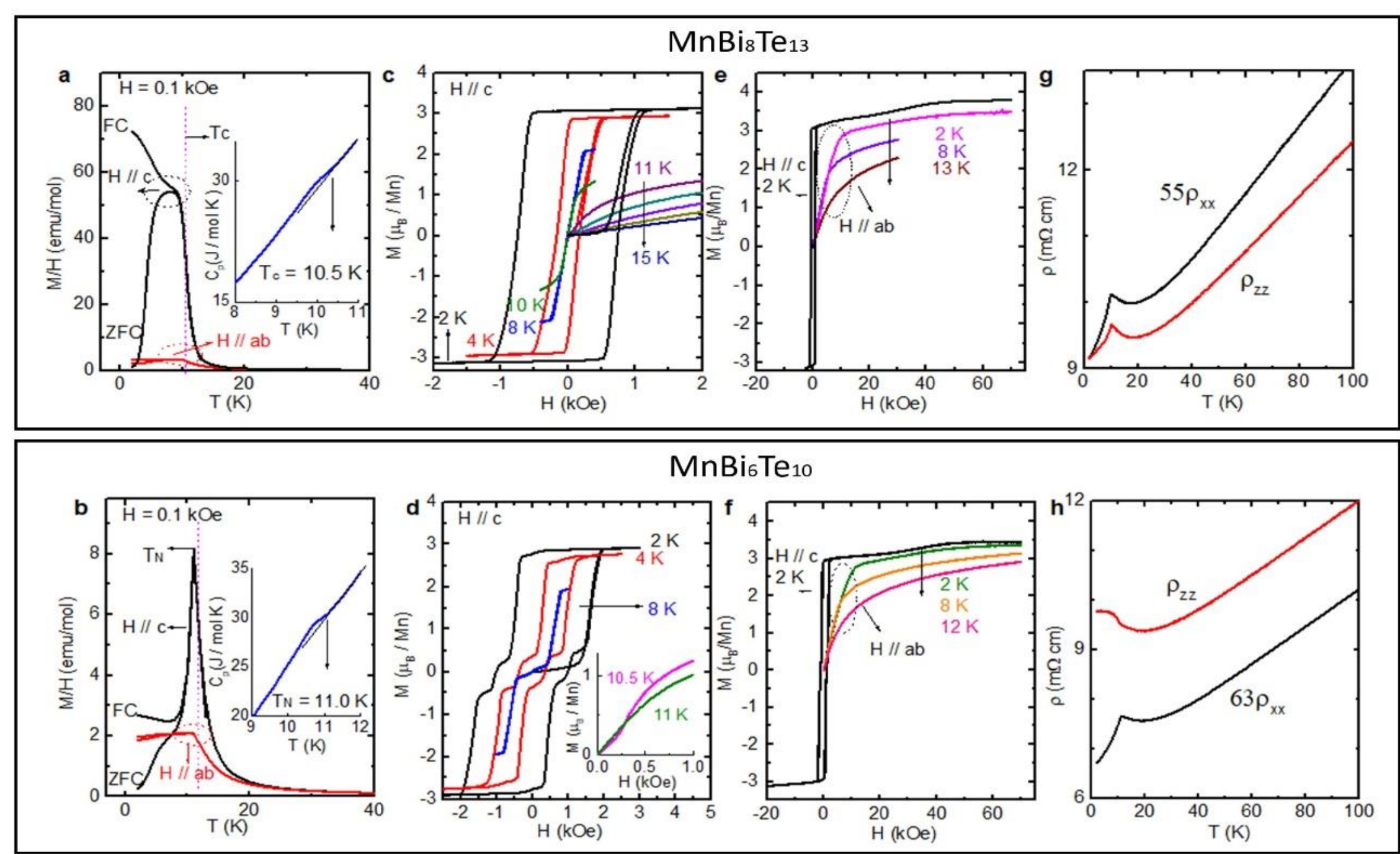

**Figure 2. Transport and thermodynamic properties of $MnBi_8Te_{13}$. As a comparison, physical properties of $MnBi_6Te_{10}$ are shown too.** **a-b** The temperature dependent susceptibility taken at $H$ = 0.1 kOe for: **a**: $MnBi_8Te_{13}$ and **b**: $MnBi_6Te_{10}$. Inset: The temperature dependent specific heat: **a**: $MnBi_8Te_{13}$ and **b**: $MnBi_6Te_{10}$. The criterion to determine $T_c$ and $T_N$ are shown in the inset. **c-d** Magnetic hysteresis loop of isothermal magnetization at low fields with $H \perp c$ for: **c**: $MnBi_8Te_{13}$ and **d**: $MnBi_6Te_{10}$. **e-f**: Magnetic hysteresis loop of isothermal magnetization up to 7 T with $H \perp c$ and $H \perp ab$: **e** $MnBi_8Te_{13}$. **f**: $MnBi_6Te_{10}$. **g-h**: Anisotropic resistivity, $\rho_{xx}$ and $\rho_{zz}$ for: **g** $MnBi_8Te_{13}$. **h**: $MnBi_6Te_{10}$.

Figure 3

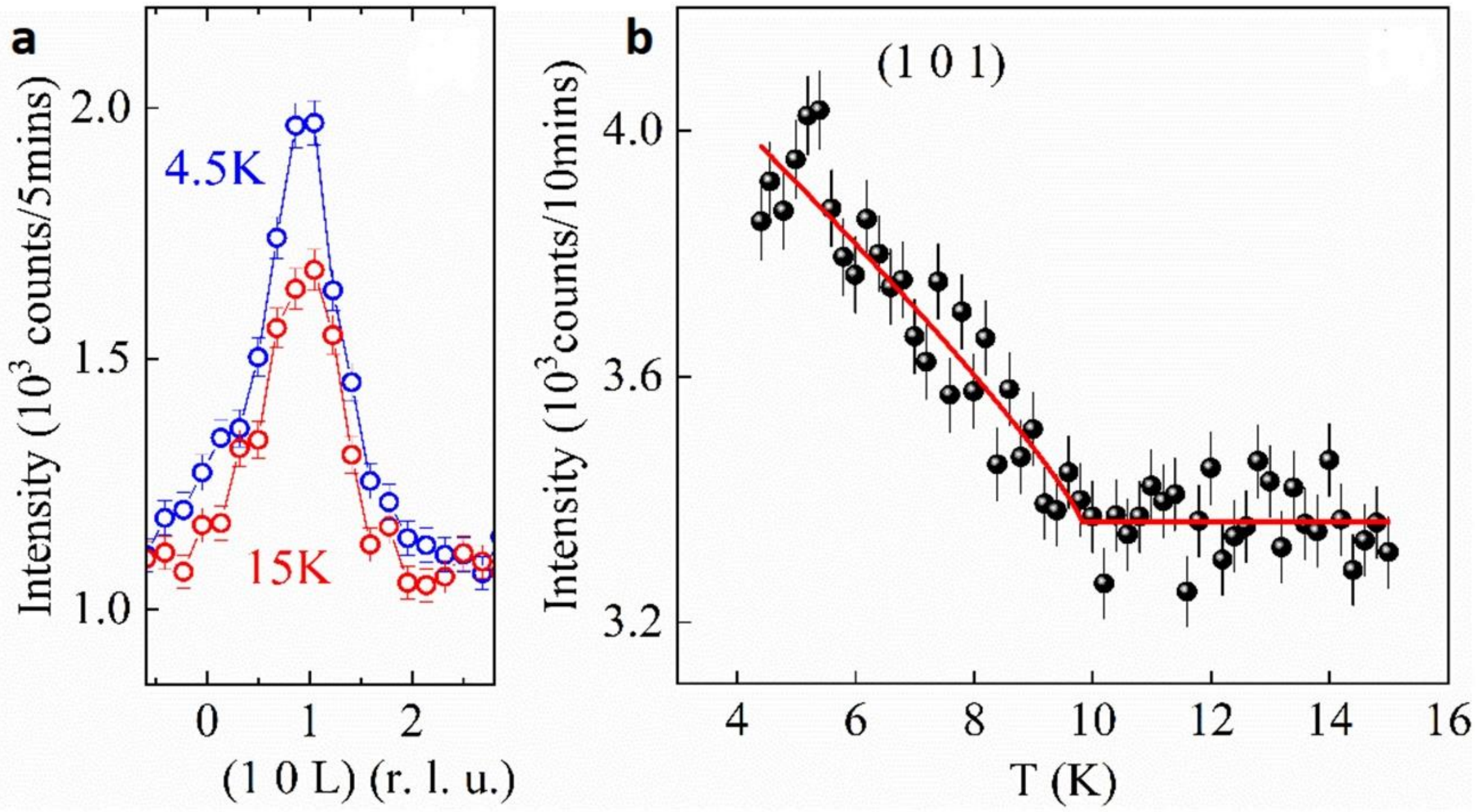

**Figure 3**. **Neutron diffraction data of $MnBi_8Te_{13}$.** **a** Rocking curve scan for Bragg peak (1 0 1) of $MnBi_8Te_{13}$ at 4.5 K and 15 K through neutron diffraction. **b** Temperature-dependent intensity of the magnetic reflection (1 0 1) of $MnBi_8Te_{13}$. The solid line stands for the power-law fit using $I = A(1 - \frac{T}{Tc})^{2\beta} + B$, where $\beta$ is the order parameter critical exponent(*39*).

Figure 4

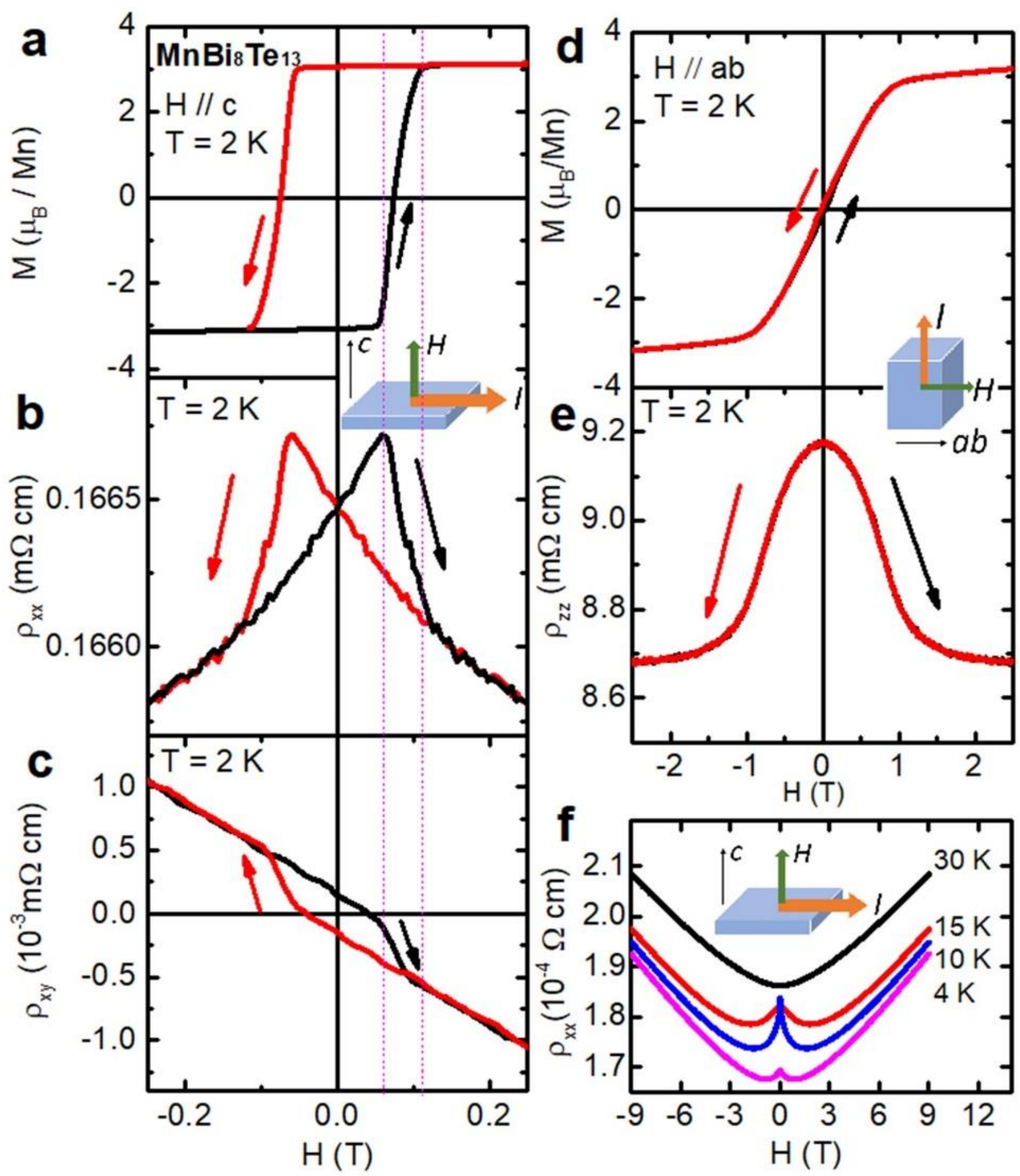

**Figure 4. Magnetotransport properties of $MnBi_8Te_{13}$, a-c** The field dependent magnetization *M* (**a**), transverse magnetoresistivity of $\rho_{xx}$ (**b**) and Hall resistivity $\rho_{xy}$ (**c**) at 2 K with *I* ‖ *ab* and *H* || *c***. d-f** The field dependent magnetization *M* (**d**), transverse magnetoresistivity of $\rho_{zz}$ (**e**) at 2 K with *I* ‖ *c* and *H* ‖ *ab*. **f** Transverse magnetoresistivity of $\rho_{xx}$ with *I* ‖ *ab* and *H* || *c* at various temperatures.

Figure 5

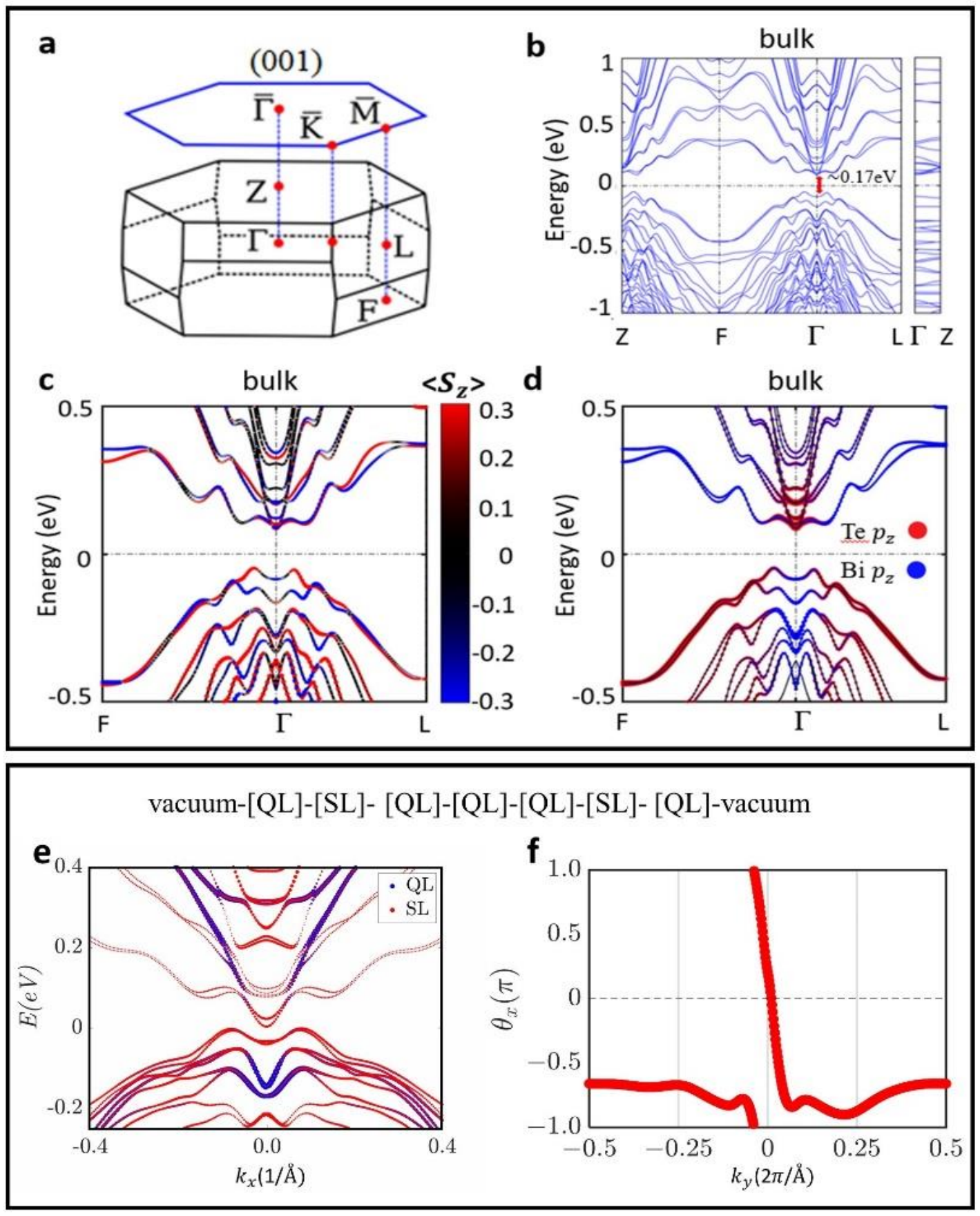

**Figure 5. DFT band structure of $MnBi_8Te_{13}$ calculated using the experimental lattice parameters with relaxed atomic positions. a-d** Bulk band structure. Bulk Brillouin zone (BZ) and the (001) surface BZ of $MnBi_8Te_{13}$ with the high symmetry points marked (**a**). Bulk band structure of $MnBi_8Te_{13}$ in the out-of-plane FM configuration with spin-orbit coupling (SOC) and correlation parameter $U$ included (**b**). Spin resolved band structure zoom in around the Γ point (**c**). Orbital resolved band structure zoom in around the Γ point (**d**). The red and blue dots indicate Te $p_z$ and Bi $p_z$ orbitals, respectively. There are clear band inversions between the Te $p_z$ and Bi $p_z$ states at the Γ point. **e-f** The DFT calculation of a 7-layered finite-size slab model corresponding to the $QL_1$ surface arrangement, i.e. with vacuum-[QL-SL-QL-QL-QL-SL-QL]-vacuum. **e.** The band structure of this slab model. The sizes of the blue and red dots represent the fraction of electronic charge residing in the topmost QL and the nearest-neighboring SL, respectively. **f.** The evolution of the sum of Wannier charge centers (WCC) along $k_y$ in the $k_z = 0$ plane. The trajectory of WCC is an open curve traversing the whole Brillouin zone once, indicating the Chern number $C = 1$ in the $k_z = 0$ plane.

Figure 6

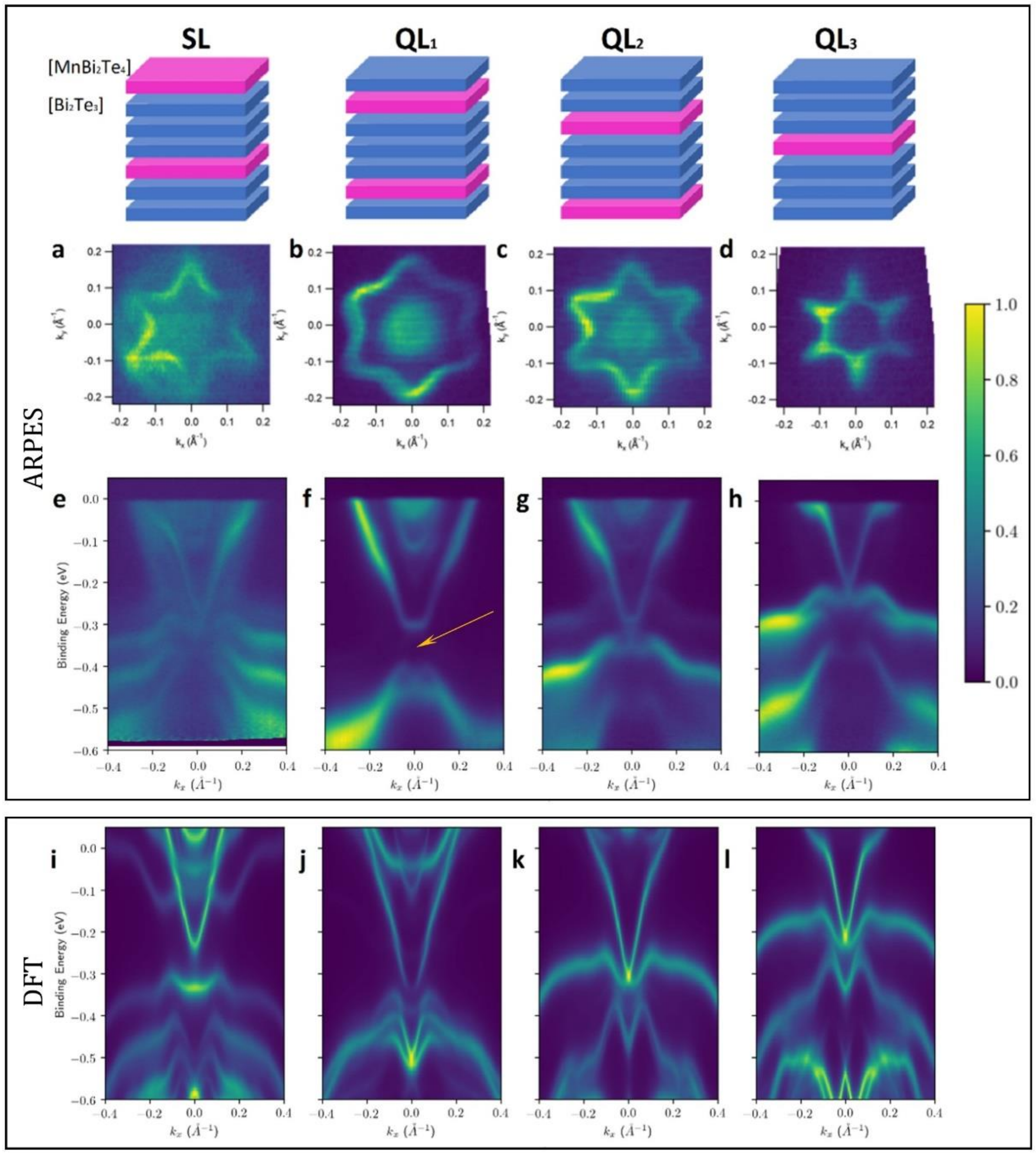

**Figure 6. The comparison of ARPES data and DFT calculation on $MnBi_8Te_{13}$. a-d** The experimental ARPES isoenergy surfaces at the Fermi level measured with 26 eV, 20×50 μm spot-sized linear horizontal light at 12 K: (**a**) SL, (**b**) $QL_1$, (**c**) $QL_2$ and (**d**) $QL_3$ termination. All Fermi surfaces show the expected six-fold symmetry with clear distinctions. **e-h** The experimental ARPES $E$-$k$ spectrum cut along the $M \rightarrow \Gamma \rightarrow M$ high symmetry direction for various terminations. A large and clean gap in the surface states of the QL1 termination is observed in panel (**f**), as highlighted by the arrow which points to the charge neutrality point suggested by DFT calculations. **i-l** The calculated DFT $E$-$k$ spectrum cut along the $M \rightarrow \Gamma \rightarrow M$ high symmetry direction for various terminations. We shift the Fermi level to match the experimentally observed Fermi level for each termination.